\documentclass[11pt,a4paper]{article}
\usepackage{moriond,epsfig,times}

\def\be{\begin{equation}}
\def\ee{\end{equation}}
\def\bea{\begin{eqnarray}}
\def\eea{\end{eqnarray}}

\begin{document}
\vspace*{2cm}
\title{LOW-MASS SUBSTELLAR CANDIDATES IN NGC\,2264}

\author{TIM KENDALL$^1$, JEROME BOUVIER$^1$, ESTELLE MORAUX$^2$ and DAVID JAMES$^3$}

\address{1. Laboratoire d'Astrophysique de Grenoble, BP 53X, 38041 Grenoble Cedex, France}

\address{2. Institute of Astronomy, University of Cambridge, Madingley Road, Cambridge, CB3 0HA, UK \\}

\address{3. Physics \& Astronomy Dept., Vanderbilt University, 1807 Station B, Nashville, TN 37235 USA\\ }

\maketitle\abstracts{NGC\,2264 is a young (3\,Myr), populous star forming region for which
our optical studies have revealed a very high density of potential brown dwarf (BD) candidates - 236 in $<$ 1\,deg$^2$ - from the substellar limit 
down to $\sim$\,20\,M$_{\rm Jup}$. Candidate BD
were first selected using wide field ($I,z$) band imaging with CFHT/12K, by reference to 
current theoretical isochrones. Subsequently,
around 50\% of the $I,z$ sample were found to have
near-infrared (2MASS) photometry, allowing further selection by
comparison with the location of DUSTY isochrones in colour-colour
diagrams involving combinations of $I$,$J$,$H$ and $K$ colours. 
After rejection of objects with only upper limits to $J$, six candidates were selected from the
$I-K,J-H$ diagram which afforded the best separation of candidate and
field objects; of these, 2 also lie close to the model predictions in the $I-J,I-K$ and $I-J,H-K$ plots. After
dereddening, all six remain probable very low-mass NGC\,2264 members, in spite of their low A$_{\rm v}$, while
a different group of objects are shown to be highly reddened background giants.   
A further three brighter (at $I$) objects selected by their $I-J,I-K$ colours, lie at the substellar limit and  are 
likely cluster objects, as are 2 intermediate mass objects selected by their
$I-K$ and $H-K$ colours. 
These objects potentially constitute a hitherto unknown population of young, low-mass BD
in this region;  
only slighty deeper observations could reveal a new laboratory for the study of near-planetary-mass objects.}

\noindent
{\small¥{\it Keywords}: stars: low mass, brown dwarfs -- infrared: stars -- surveys -- Galaxy: open clusters and associations 



\section{Introduction}

Within the last few years the observational study of substellar objects (M\,$<$0.072\,M$_{\odot}$) has undergone
spectacular and rapid development, opening up new perspectives on the formation of such objects within molecular clouds.
Large numbers of BD have now been found in star-forming regions, young clusters and the 
field \cite{bej01}$^,$ \cite{mor03}$^,$  \cite{cha02}$^,$ \cite{cha03} and physical models of the atmospheres of BD 
constructed \cite{bar03}$^,$ \cite{cha00}, opening up the possibility of confrontation between observations and 
theoretical predictions of the physical properties of BD. 
However, the core issues of the form of the substellar initial mass function (IMF) 
and its dependence on
environment remain to be addressed. The discovery of BD in widely differing environments does suggest their formation is
directly linked to the star formation process, and early estimates of the substellar IMF in the solar neighbourhood 
have indicated that BD are nearly as numerous as stars. Within this broad framework, two competing scenarios have been
put forward for BD formation; the first simply that they form as stars do, i.e. by the gravitational collapse of low mass
molecular cloud cores \cite{pad02}; the second postulates the dynamical ejection of the lowest mass protostars, leading to
BD formation since the ejected fragments are unable to futher accrete \cite{rei01}. To distinguish between these
possibilities and to obtain unbiased estimates of the substellar IMF, observations of statistically complete, homogeneous
populations of BD are needed in a wide variety of environments with different ages. Such a deep, wide-field 
imaging survey has been performed in the $I,z$ bands with CFHT/12K/MegaCam: ``Brown 
dwarfs and the substellar mass function: Clues to the star formation process'', initiated by J. Bouvier and collaborators. 

The study of NGC\,2264 introduced here is only a small part (0.6\,deg$^2$) of the whole survey, 
which, for the first time, has covered
significant areas (70\,deg$^2$) of star-forming environments, pre-main sequence clusters and older open clusters (ages
1--600\,Myr) with the aim of reaching $I$\,$\sim$\,24 which corresponds to 25\,M$_{\rm Jup}$ at the distance of Taurus (140\,pc)
for age 10\,Myr and extinction A$_{\rm v}$\,=\,10 mag; in all of the targeted regions, the limiting mass is in the
range 10--40\,M$_{\rm Jup}$ and NGC\,2264, because of its youth and relative proximity 
(3\,Myr, 770\,pc \cite{lam04}$^,$ \cite{reb02}), is probed to the lower
end of this mass range with the current observations.

\section{Young brown dwarfs in star-forming regions (SFRs)}

Young BD (with ages 1--3\,Myr) are now being uncovered (for a recent compilation see \cite{bas00})
but the census remains very incomplete to date. The CFHT project as a whole has now discovered a substantial
sample of candidate young substellar objects - several hundred to date -  which can be studied statistically
with confidence and which can provide targets for observations in other wavelength regimes
(eg. with the Spitzer Telescope). The youth of such objects  ensures that their population has not yet
suffered from important dynamical (and stellar) evolution, other than dynamical effects 
potentially associated with their formation. This is a point worth underlining, since observations of
young SFRs have the potential to distinguish the two BD formation mechanisms outlined in Sect.\,1
by their resultant spatial distributions. If BD always form exactly as stars do, one would expect them to
trace the regions of the highest density of low-mass stars. However, in any dynamical ejection scenario,
a deficit of BD may be expected in the central regions of young star-forming clusters; their initial
velocities ($\sim$\,1\,km\,s$^{-1}$) \cite{ste03} moving them away from their birth sites at
$\sim$\,0.3$^{\circ}$/Myr (at 140\,pc). Such a deficit would therefore be observable in all nearby
young SFRs. 

Comparison with young SFRs with a range of environmental conditions
(from the low density regions such as Taurus to regions where stars are forming in very massive 
clusters, such as Upper Scorpius) also permits investigation of the sensitivity of the low-mass
end of the IMF to local conditions. Sufficiently large coeval populations are not yet 
available to perform such studies with confidence, but recent studies are beginning to address 
such questions. 
As an example, 30 substellar candidates have been identified in Taurus in a 3.6\,deg$^2$ region using CFHT/12K
data which reached to $I$\,=\,23.5. Spectroscopic follow-up led to the identification of
four BD in Taurus \cite{mar01} with spectral types later than M7. In Taurus, $\sim$10 BD are
now known \cite{bri02}, and appear to be spatially correlated with regions of highest  stellar
density. In Upper Scorpius, initial studies \cite{mar04} have found 18 candidate BD of which 5
show signs of ongoing accretion. We will compare our initial findings on NGC\,2264 in Sect.\,4;
here we will note that the much larger BD populations being uncovered by CFHT/12K, when fully
analysed, promise a more robust statistical treatment of these issues.   

\begin{figure}
\begin{minipage}{9cm}
\hspace*{3cm}
\epsfig{figure=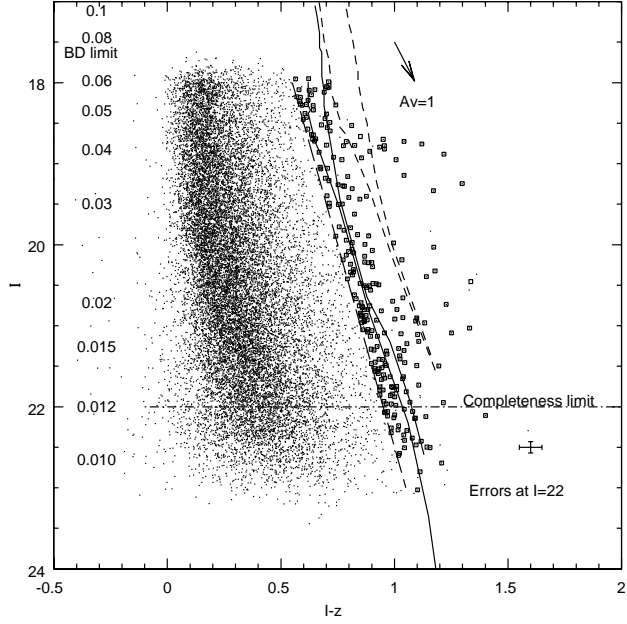,height=9cm}
\end{minipage}
\caption{($I$,$I-z$) 
colour-magnitude diagram built using long (360\,sec) exposures of NGC\,2264.
The solid isochrones are DUSTY models for an ages 2 and 5\,Myr; the dashed isochrones
are the NextGen models at the same ages. All isochrones are for a distance of 770\,pc. 236
 BD candidates (squares)
were selected to be redward of the sloping straight line (long dash). Small dots
 redward of this
line were rejected from the candidacy on visual inspection. The mass scale 
(in M$_{\odot}$) is for the 2\,Myr models; it and the estimated completeness limit are indicated.}
\end{figure} 

\section{Candidate NGC\,2264 brown dwarfs: Optical and near-infrared colour selection}

The pre-reduction and analysis of these date have been performed using CFHT Elixir pipelines\,\footnote{www.cfht.hawaii.edu}
and innovative point-spread function fitting techniques developed by E. Bertin at the Institut d'Astrophysique 
de Paris\,\footnote{www.terapix.iap.fr}. 
Full
details of these methods will be published in a later paper. A first selection of candidates was made by choosing those
redder than the long dashed line in Fig.\,1., 
which shows optically selected candidates in NGC\,2264 between the substellar limit and $\sim$\,10M$_{\rm Jup}$, 
by comparison with the state-of-the-art DUSTY models, specially created by Baraffe et al. to take account of the 
CFHT $I,z$ filter responses (see \cite{bar03} and references therein for full details of these models).  
All 236 such $I,z$ candidates
(squares in Fig.\,1) have been visually
inspected on both $I$ and $z$ frames and all found to be stellar in nature, i.e. there are no artifacts
due to nebulosity, field edges, bad columns, bright stars etc. As can be seen in Fig.\,1, we estimate our
data are complete to $I$\,$\sim\,$22, or 12\,M$_{\rm Jup}$ for age 2\,Myr.
Cross-correlation with the 2MASS all sky release data has yielded 101 counterparts to the 
$I,z$ candidates with $JHK$
magnitudes accurate to $\pm$\,0.3 magnitudes or better (a further 29 have only upper limits to $J$ (and for 5, $K$). 
These counterparts are plotted in Fig.\,2(a)
as large squares, together with the complete set of 2MASS data within the optically surveyed area (crosses), in the
$I-K,J-H$ colour-colour diagram. 
It is clear that this diagram provides an excellent separation between
candidates and field objects. A  number of candidates are extremely red,
with $I-K\sim$\,6 and $I-J\sim$\,3--4; the latter colour being typical of field objects with late M or
L spectral types, e.g.\,\cite{ken04}.

\begin{figure}
\begin{minipage}{9cm}
\epsfig{figure=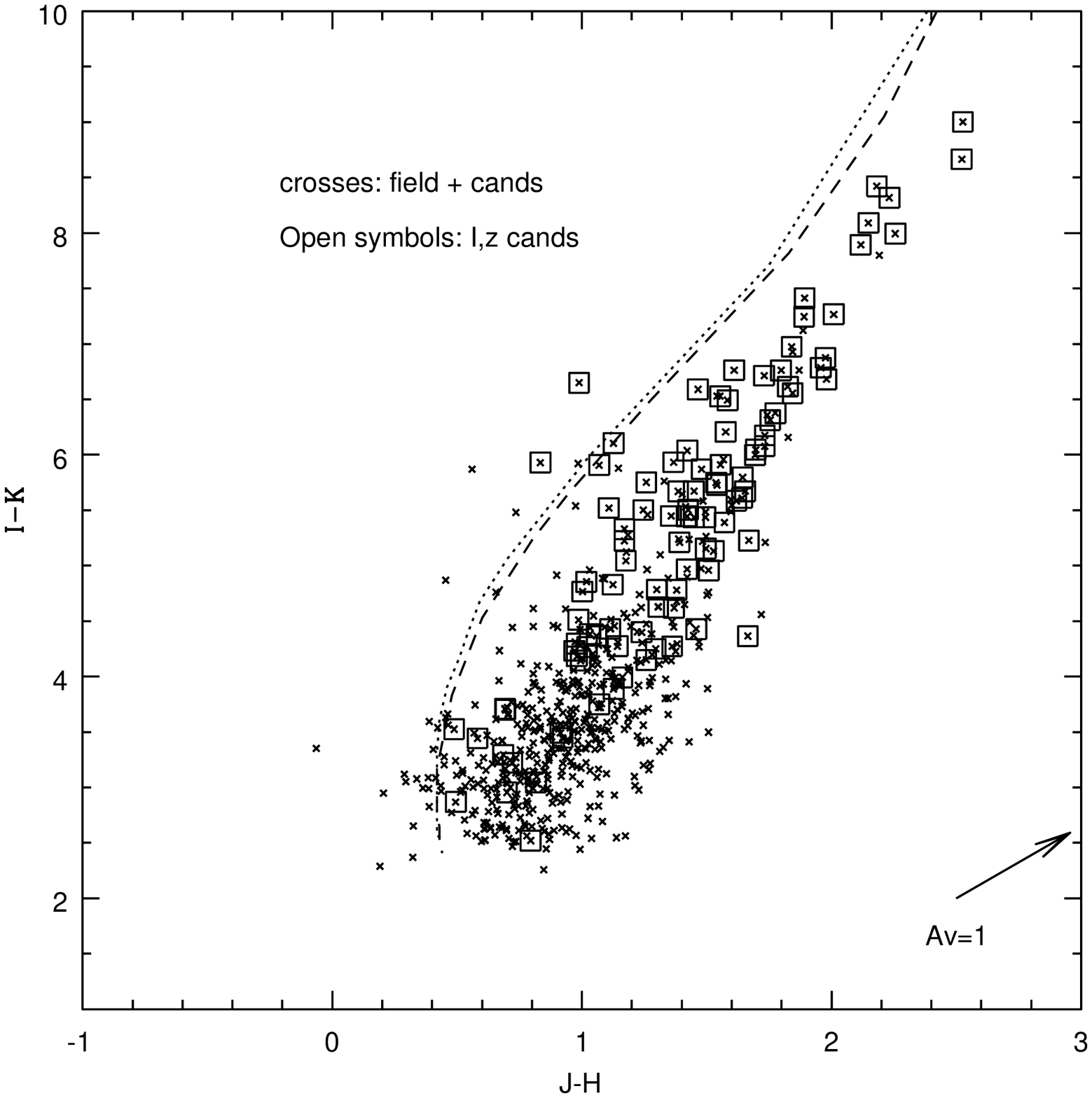,height=8cm}
\end{minipage}
\hspace*{-1cm}
\begin{minipage}{9cm}
\epsfig{figure=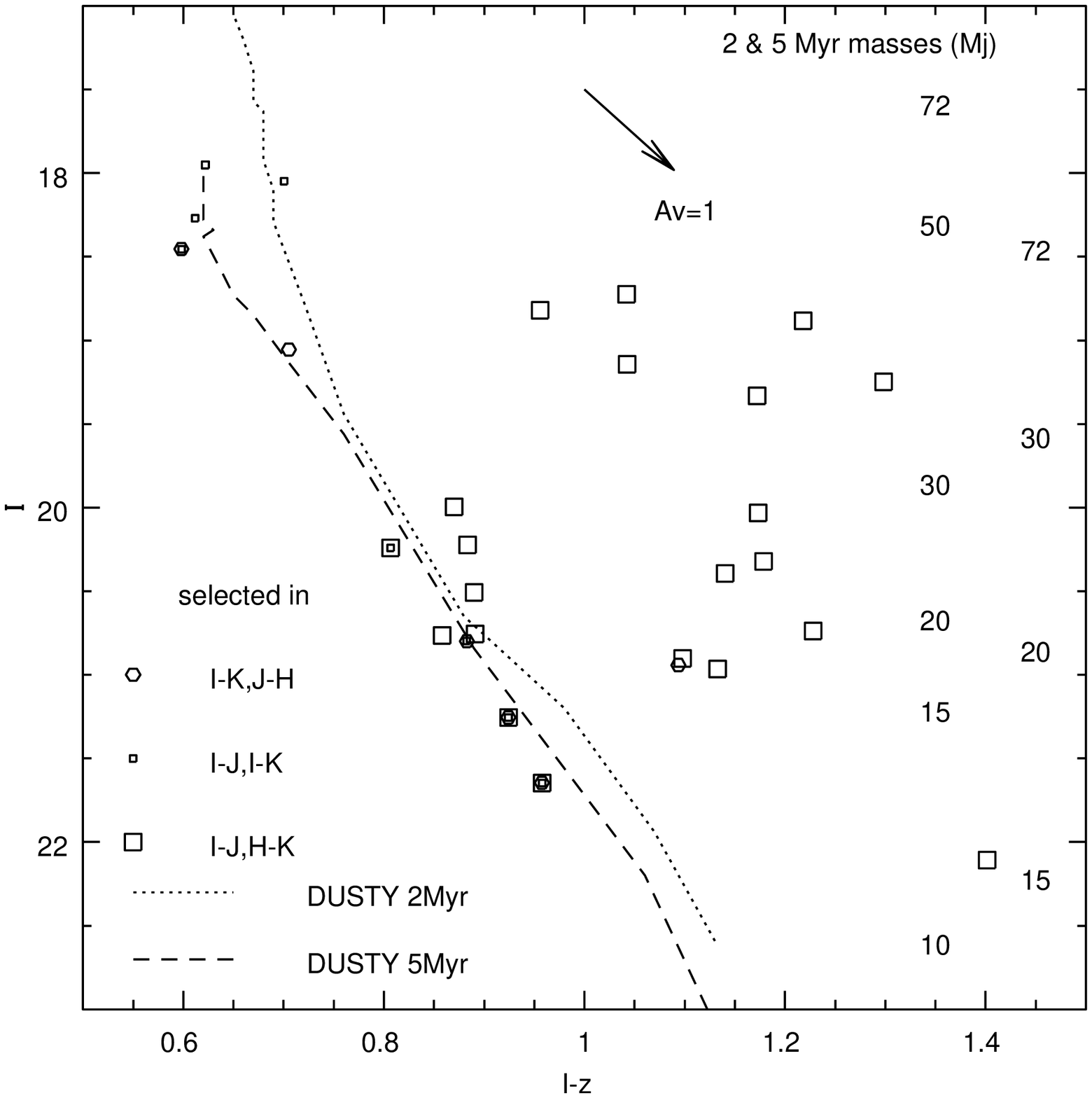,height=8cm}
\end{minipage}
\caption{(a, Left): All $I,z$ candidates with 2MASS (squares) and all 2MASS points (field and 
candidates, crosses)
in the $I-K,J-H$ colour-colour diagram. Note the separation of the two 
populations and the
reddening vector. The dashed curve is the 5\,Myr DUSTY isochrone; the dotted, 
1\,Myr. Figure 2: (b, Right):  Final selection of candidates from colour indices in $I,J,
H,K$, plotted in the ($I,I-z$) diagram, and excluding objects with only $JK$ upper limits. 
It is clear that the the $H-K$ colour is least
efficient at selecting objects near the model loci. 
DUSTY isochrones only  are plotted as in Fig.\,1. Candidates were selected from their locations in the $I-K,J-H$ diagram {\it and} two other
colour combinations (see key). Typical errors in $I$ and $I-z$ are small, $\sim$\,0.1\,mag. (see Fig.\,1).}
\end{figure}

\begin{figure}
\begin{minipage}{9cm}
\epsfig{figure=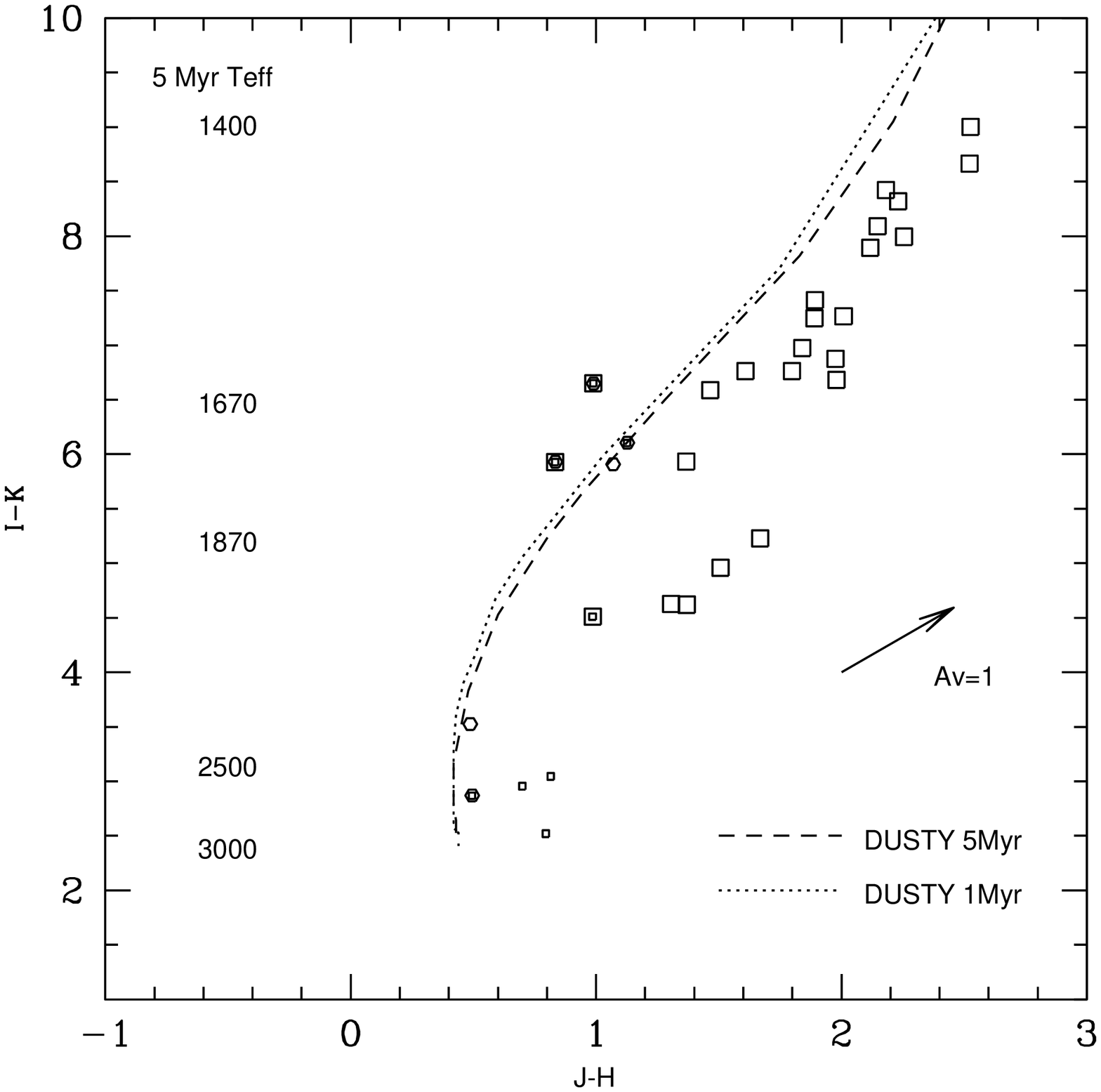,height=8cm}
\end{minipage}
\hspace*{-1cm}
\begin{minipage}{9cm}
\epsfig{figure=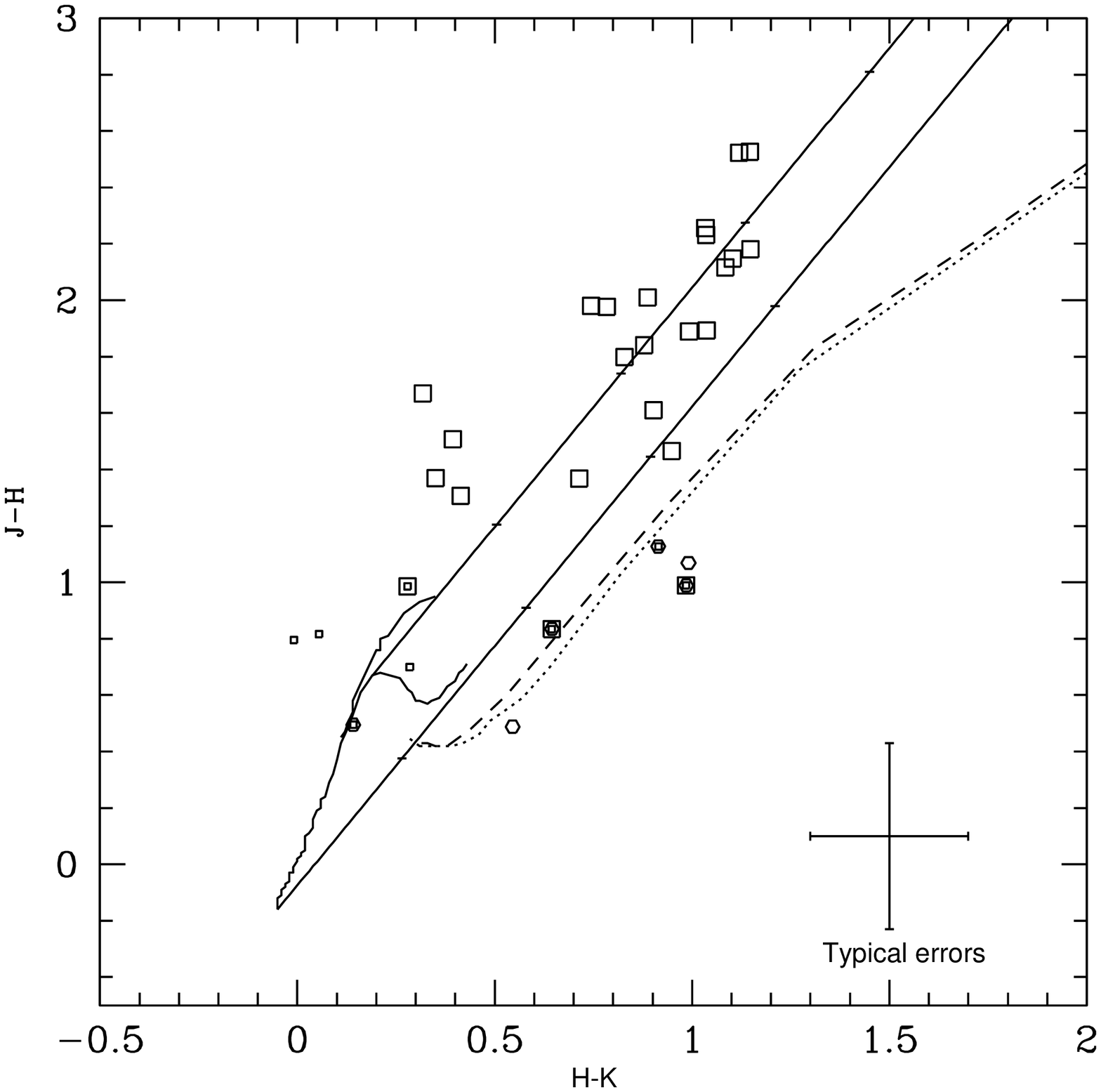,height=8cm}
\end{minipage}
\caption{(a, Left): Final census of  near-infrared selected objects plotted in the ${I-K,J-H}$ diagram, as an aid to visualizing the
selection space. 
Note the T$_{\rm eff}$ scale based on $I-K$ colour. Figure 3: (b, Right): Location in the $JHK$ diagram. The reddening band (RB) is
shown (each horizontal tickmark represents A$_{\rm v}$\,=\,5), together with loci for dwarfs and giants.
Objects selected only in 
$I-J,H-K$ are clearly more likely to be reddened contaminants; the most promising candidates
lie away from the RB and close to the DUSTY isochrones in this plot. Some may have small $K$-band excesses. 1 and 5\,Myr DUSTY isochrones
are plotted as for Fig.\,3(a). Typical 2MASS errors are given; the error in $I-K$ is somewhat smaller than in $J-H$ or $H-K$, 
as it is dominated by that in $K$ and the
error in $I$ from our 12K data is comparatively small.}
\end{figure}

To further refine our sample, we have followed similar methods
to those of \cite{bri02}$^,$ \cite{luh03} in Taurus, using combinations of $I,J,H,K$ colour-colour diagrams
to select those optical candidates which lie on or close to the model isochrones in these diagrams. For the similarly young and extinction affected 
NGC\,2264 region, our finding that the $I-K,J-H$
diagram yields the best separation between field 2MASS objects and candidate BD is in agreement with these authors. 
In practice,
we have chosen candidates lying close to the isochrone in Fig.\,2(a), after discarding $I,z$ candidates with only upper limits to $J$ or $K$. Fig.\,2(b) shows
objects selected in various colour combinations (see key) re-plotted in the
$I,I-z$ diagram, and, in Fig.\,3(a), the $I-K,J-H$ diagram.  
It can be clearly seen that objects selected by $I-J,H-K$ colours {\it only} (open squares in Figs.\,2(b) and 3(a)) do not as a group
fall close to the cluster isochrones in $I,I-z$ and these objects are almost certainly reddened background giants.
Indeed, in a conventional $JHK$ colour-colour
diagram (Fig.\,3(b)) many objects {\it only} selected by their $I-J,H-K$ colours lie within the reddening band
However, candidates selected in the $I-J,I-K$ and, especially, $I-K,J-H$ diagrams.
tend to lie away from the reddening band, and closer to the cluster isochrone. They can therefore can be considered much more likely
to be substellar NGC\,2264 members. Such candidates are plotted by small open squares and open hexagons
in Figs.\, 2(b) and 3. 

These arguments are backed up by simple dereddening in $J-H$ where all objects have been taken to the low-mass tip of the dwarf
sequence in Fig.\,3(b), and subsequently replotted in the $I,I-z$ diagram (Fig.\,4(a)). To deredden $z$, we interpolate the interstellar reddening laws
of \cite{rl85} to find A$_{\rm z}$/A$_{\rm v}$\,=\,0.406, taking the optimum sensitivity of the $z$ filter at 9800\AA, as given by the CFHT website. 
As a group, the $I-J,H-K$
selected objects are proven to be highly reddened with A$_{\rm v}$ ranging up to $\sim$\,17. Only one such object might
be considered a member by reference to its position near the tip of the 2\,Myr isochrone in $I,I-z$; one further 
candidate is simultaneously selected in $I-J,I-K$ and has a predicted mass near 40\,M$_{\rm Jup}$. Six objects, all part
of the $I-K,J-H$ selection, lie close to the isochrones with masses ranging down to 20\,M$_{\rm Jup}$, and three more chosen
by their $I-J,I-K$ colours only are clustered near the model predictions very close to the BD limit. Therefore, the
total number of probable substellar members found in NGC\,2264 and reported here is eleven. 

\begin{figure}
\begin{minipage}{9cm}
\epsfig{figure=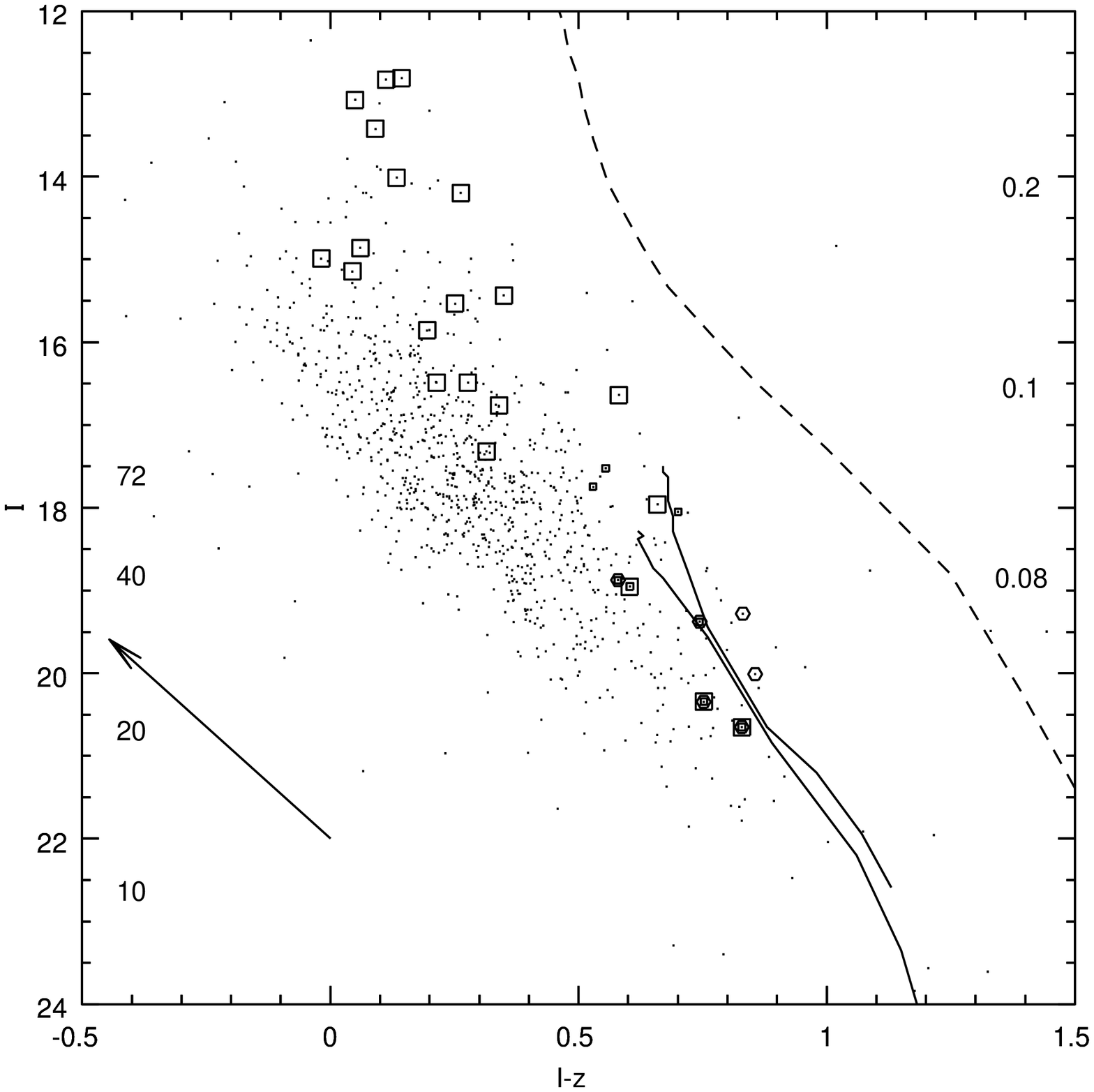,height=8cm}
\end{minipage}
\hspace*{-1cm}
\begin{minipage}{9cm}
\epsfig{figure=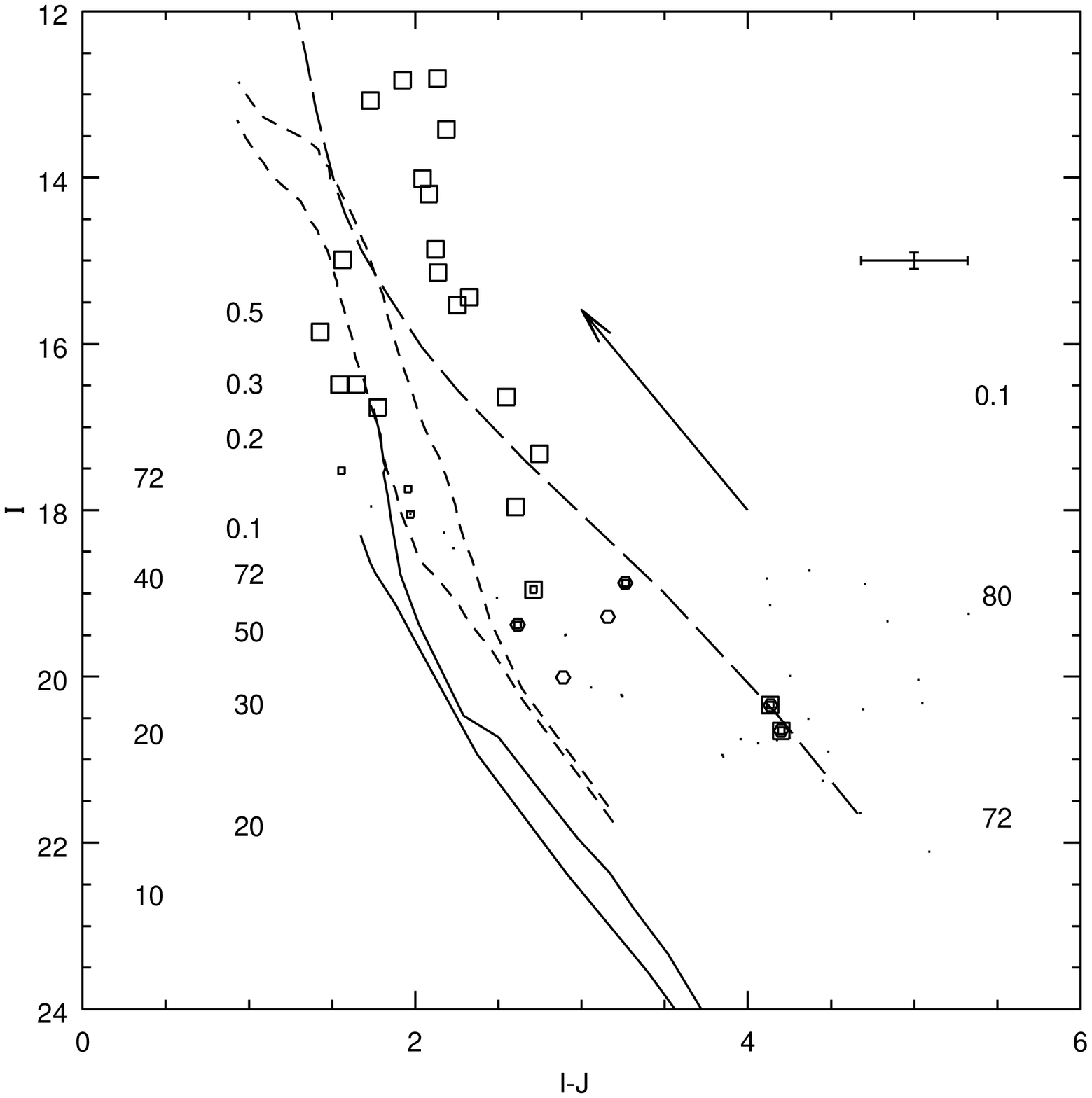,height=8cm}
\end{minipage}
\caption{(a, Left): Locations of objects in $I,I-z$ after simple dereddening described in the text. Most $H-K$ selected
objects (open squares) are clearly in the background. The dots are a locus of {\it all} 2MASS points in the region, similarly dereddened;
the locus of all field points. The best cluster candidates selected by the methods described here lie noticeably away from this locus.
The arrow represents the dereddening vector for A$_{\rm v}$\,=\,5. 
2 and 5\,Myr DUSTY isochrones are plotted as solid curves for which the mass scale at left is for 2\,Myr; 
the dash is the location of a 5\,Gyr isochrone,
representing the field, at 60\,pc.  (b, Right): As Fig.\,4(a) but for $I,I-J$. DUSTY isochrones are solid curves, short dashes NextGen, for 1 and 5\,Myr ages. 
The long dash is the field isochrone for 60\,pc. Dots are locations before dereddening. 
Mass scales are for the 5\,Myr DUSTY (extreme left), 2\,Myr NextGen (left) and the field model (right). 
The possibility that the two points lying on it at $\sim$\,75\,M$_{\rm Jup}$ are {\it foreground field objects} is discussed in the text. A typical errorbar
is shown.}
\end{figure}


\section{Discussion and concluding remarks}

The first and most important point concerns the masses of our candidates, derived from their $I$ magnitudes after dereddening, 
by comparison with theoretical models for 2 and 5\,Myr plotted for a distance modulus 9.4. For the 11 most probable cluster members shown 
in Fig.\,4(a) and discussed above, predicted masses range between the substellar limit and close to 20\,M$_{\rm Jup}$.  

However we caution that the photometric methods used in this work do not rule out that some of the candidate BD may be background
giant or faint foreground contaminants. This is an important point, in the light of the obvious
existence of strong and variable extinction intrinsic to the NGC\,2264 region  itself as discussed at length by \cite{reb02} (and references therein), and in the
foreground, since the region lies close to the galactic plane. Clearly, our methods can distinguish background giants. However, \cite{reb02} 
draw attention to the surprising fact that a number of {\it known}
NGC\,2264 members exhibit only the reddening expected in this line of sight at $b$\,=\,2$^{\circ}$, E$_{\rm B-V}$\,$\sim$\,0.5. Presumably, these
members must lie on the near edge of the cloud and be unaffected by reddening intrinsic to the cluster nebulosity itself. The eleven objects
suggested here as substellar members have A$_{\rm v}$ typically in the range 0--3 and {\it not more than 4}, as apart from the objects we have 
clearly identified as background giants suffering $\sim$\,15\,magnitudes 
of extinction. Their A$_{\rm v}$ are therefore perfectly consistent with cluster
membership, although they may be preferentially located on the near edge of the cloud. 

This discussion brings to the point the question of contamination by {\it foreground} objects. It is seen from inspection of the $I,I-J$
diagram in Fig.\,4(b) that the candidates are spread in colour between the cluster models and an isochrone plotted for age 5\,Gyr (i.e.
representing the {\it field} population) at a distance chosen to be 60\,pc to bisect the two candidates selected in all colours and 
shown as triply overplotted points. For reasons which we defer to a future paper, the question of foreground contamination is not raised
by the location of the same isochrone in $I,I-z$ (Fig.\,4(a)). However, the status of these 2 objects in particular, as questioned by their
dereddened $I-J$ colour, can be further investigated. These objects have A$_{\rm v}$ of 1.25 and 2.7, as determined by our method. At first
sight, it would appear that it is very unlikely that a foreground object at 60\,pc would suffer such extinction, supporting the claim that
they are true NGC\,2264 members with masses not near the BD limit but instead much lower. Indeed, for one of these objects the 2MASS $J$-magnitude
is flagged ``D'' and its $I-J$ colour might be bluer (nearer the cluster isochrone) than plotted; the other is flagged ``C'' and may also
be bluer by at least the typical error shown in Fig.\,4(b).   

Further general arguments favourable to the identification of these eleven candidates as NGC\,2264 members are provided by the observation of
clear separation of $I,z$ selected objects from the field in the $I-K,J-H$ diagram and by the classical $J-H,H-K$ diagram (Fig.\,3(b)), in which
the location of
DUSTY isochrones has {\it not} been employed for candidate selection. It can be
readily seen that, even before any dereddening, the best candidates as a group here lie away from the reddening band, while other initially
plausible candidates, selected using their $I-J$ and $H-K$ colours only, are more scattered over the diagram, often
lying within the reddening band. A similar separation identifies background giants in the dereddened $I,I-z$ plot. Such observations 
further strengthen the effectiveness of our photometric
selection methods, with $I-K,J-H$ selected candidates preferentially located closest to the DUSTY isochrones
in all colour-colour and colour-magnitude diagrams. The location of a few candidates in $JHK$ also raises the possibility that they have $K$-band excesses,
which might indicate ongoing accretion, and can be tested first by photometric observations in the thermal infrared with Spitzer. 
Current studies do not easily show whether the presence of disks might be still expected around young BD at the age of NGC\,2264, and our candidates
provide a good opportunity to constrain disk ages in this way. 

It is clear however that near-infrared spectroscopy is also required to confirm candidates
on the basis of their T$_{\rm eff}$ and surface gravities, which will be significantly less than foreground, evolved field
objects owing to the young age of the region. 

In conclusion, there is a strong likelihood that our preliminary work has indeed uncovered a
small initial sample of low mass brown dwarfs, some of which may have masses only twice the
deuterium burning limit. Deeper near-infrared photometry than given by 2MASS is likely to identify more of our $I,z$ candidates
as near-planetary mass BD. If so, then NGC\,2264 will prove to be a new astrophysical laboratory, intermediate in stellar density between
Orion and Taurus,
where the formation mechanisms of such objects in star-forming environments can be investigated.

\section*{Acknowledgments}

TRK and DJJ acknowledge support from the 5th Framework European Union Research Training Network ``The Formation and
Evolution of Young Stellar Clusters'' (RTN1-1999-00436) and TRK from the French {\it Minist\`{e}re du Recherche}.
EM acknowledges support from PPARC. 


\end{document}